\documentclass{webofc}
\usepackage[varg]{txfonts}   
\begin{document}
\title{Tau decays into two mesons: an overview}
%
%

\author{\firstname{Sergi} \lastname{Gonz\`{a}lez-Sol\'{i}s}\inst{1}\fnsep\thanks{\email{sgonzal@iu.edu}} 
}

\institute{Department of Physics, Indiana University, Bloomington, IN 47405, USA\newline
Center for Exploration of Energy and Matter, Indiana University, Bloomington, IN 47408, USA
          }

\abstract{%
We review the state-of-the-art theoretical analyses of tau decays into a pair of mesons and a neutrino.
The participant vector and scalar form factors, $f_{+}(s)$ and $f_{0}(s)$, are described in the frame of Chiral Perturbation Theory with resonances supplemented by dispersion relations, and the physical parameters of the intermediate resonances produced in the decay are extracted through the pole position of $f_{+,0}(s)$ in the complex plane. 
As a side result, we also determine the low-energy observables associated to the form factors.
We hope our study to be of interest for present and future experimental analyses of these decays.

}
\maketitle
\section{Introduction}
\label{intro}
The tau is the only lepton heavy enough ($m_{\tau}\sim1.8$ GeV) to decay into hadrons.
At the exclusive level, the hadronic partial width ($\sim65\%$) is the sum of the tau partial width to strange ($\sim3\%$) and to non-strange ($\sim62\%$) hadronic final states, and provides an advantageous laboratory to investigate the non-perturbative regime of QCD under rather clean conditions that is useful to understand the hadronization of QCD currents, to study form factors and to extract resonance parameters. 
While the non-strange decays are largely dominated by the $\pi^{-}\pi^{0}$ mode which, in turn, constitutes the main decay channel of the $\tau$ with an absolute branching ratio of $\sim25\%$, the strange hadronic final states are suppressed with respect to the non-strange ones mainly due to the following two reasons: $i)$ the mass of the strange quark is larger than the mass of the up and down quarks thus yielding to a phase-space suppression; $ii)$ strange decays are Cabibbo suppressed since the $|V_{us}|$ element of the CKM matrix enters the transition instead than $|V_{ud}|$.
The dominant strangeness-changing $\tau$ decays are into $K\pi$ meson systems which adds up to $\sim42\%$ of the strange spectral function. 
However, in order to increase the knowledge of the strange spectral function, the $\tau^{-}\to K^{-}\eta^{(\prime)}\nu_{\tau}$ decays are important.


In this letter, we provide a brief overview of the main results we have obtained in our series of dedicated analyses of two meson tau decays based on the framework of Resonance Chiral Theory supplemented by dispersion relations i.e. $\tau^{-}\to\pi^{-}\pi^{0}\nu_{\tau}$ and $\tau^{-}\to K^{-}K_{S}\nu_{\tau}$ \cite{Gonzalez-Solis:2019iod}, $\tau^{-}\to K_{S}\pi^{-}\nu_{\tau}$ and $\tau^{-}\to K^{-}\eta^{(\prime)}\nu_{\tau}$ \cite{Escribano:2014joa,Escribano:2013bca}, and $\tau^{-}\to\pi^{-}\eta^{(\prime)}\nu_{\tau}$ \cite{Escribano:2016ntp}.


\section{Theoretical framework}
\label{sec-1}
\begin{figure}[h]
\centering
\includegraphics[width=4cm,clip]{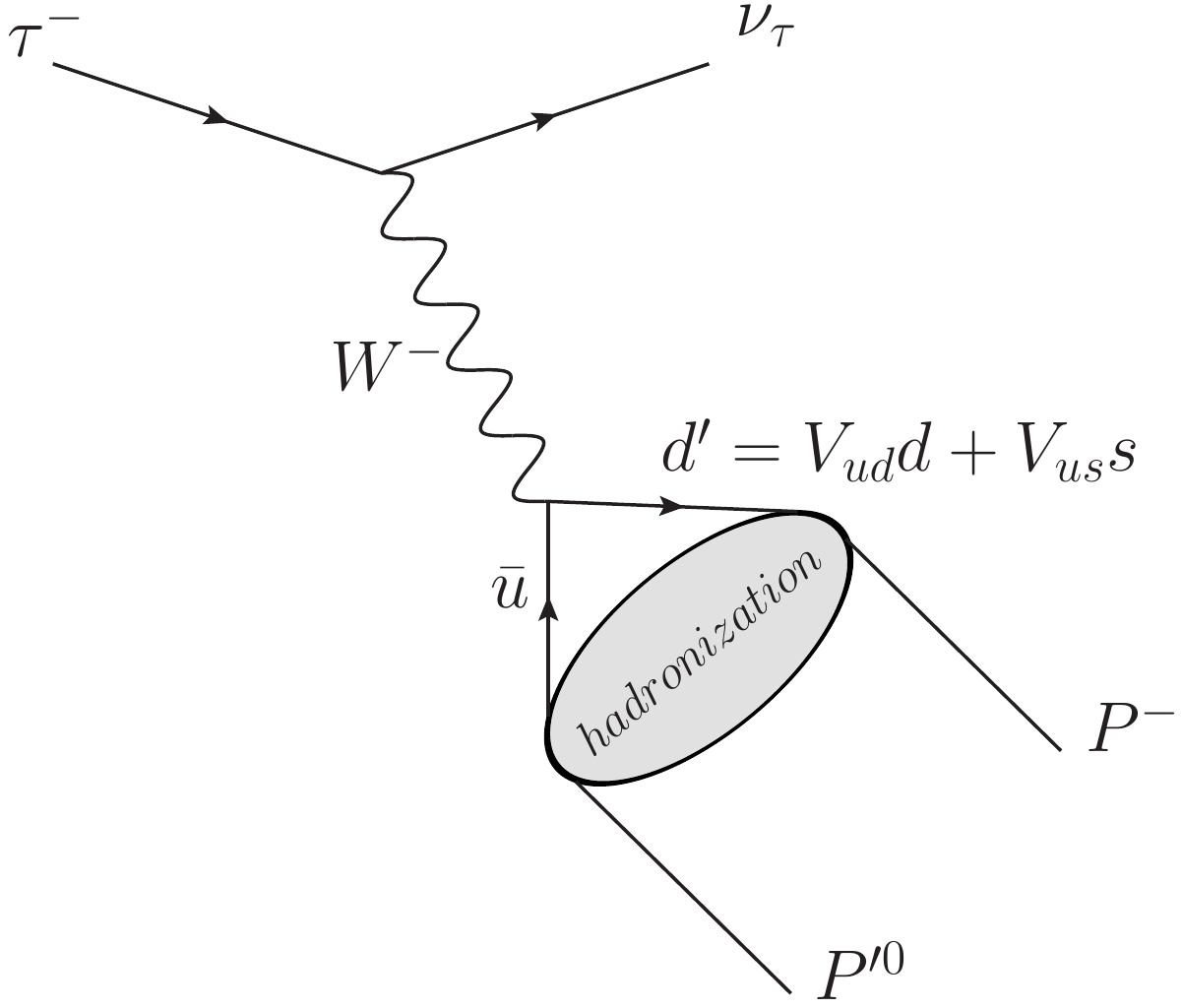}
\caption{Schematic picture of a tau decaying into two mesons.}
\label{fig-1}       
\end{figure}

Tau decays into two mesons proceeds through the exchange of $W^{\pm}$ gauge bosons which couple the tau and the generated neutrino with the quark-antiquark pair that subsequently hadronizes into a pair of mesons $P^{-}P^{0}$ (see Fig.\,\ref{fig-1} for a schematic representation).
The corresponding amplitude can be expressed as an electroweak part times an hadronic matrix element
\begin{equation}
\mathcal{M}\left(\tau^{-}\to P^{-}P^{\prime 0}\nu_{\tau}\right)=\frac{G_{F}}{\sqrt{2}}\bar{u}(p_{\nu_{\tau}})\gamma^{\mu}(1-\gamma^{5})u(p_{\tau})\langle P^{-}P^{\prime 0}|d^{\prime}\gamma^{\mu}u|0\rangle\,,
\label{amplitude}
\end{equation}
where $d^{\prime}=V_{ud}^{*}\bar{d}+V_{us}^{*}\bar{s}$.
In Eq.\,(\ref{amplitude}), we have not considered the gauge boson propagator, since the explored energy region ($\sqrt{s}<m_{\tau}$) is much lighter than the $W^{\pm}$ mass ($M_{W^{\pm}}\sim80$ GeV), but rather its expansion and used the well-known relation $G_{F}/\sqrt{2}=g^{2}/8M_{W}^{2}$.
The hadronic matrix element encodes the unknown QCD dynamics and it is given by
\begin{equation}
\langle P^{-}P^{\prime 0}|d^{\prime}\gamma^{\mu}u|0\rangle=\mathcal{C}_{P^{-}P^{\prime 0}}\Bigg\lbrace \left(p_{-}-p_{0}-\frac{\Delta_{P^{-}P^{\prime 0}}}{s}q\right)^{\mu}f_{+}^{P^{-}P^{\prime 0}}(s)+\frac{\Delta_{P^{-}P^{\prime 0}}}{s}q^{\mu}f_{0}^{P^{-}P^{\prime 0}}(s)\Bigg\rbrace\,,
\label{matrixelementff}
\end{equation} 
where $\mathcal{C}_{P^{-}P^{\prime 0}}$ are Clebsch-Gordon coefficients, $p_{-}^{\mu}$ and $p_{0}^{\mu}$ are the momenta of the charged and neutral pseudoscalars, respectively, $q^{\mu}=(p_{-}+p_{0})^{\mu}$ is the momentum transfer and $s=q^{2}$.
In Eq.\,(\ref{matrixelementff}), $f_{0}^{P^{-}P^{\prime0}}(s)$ corresponds to the $S$-wave projection of the state $\langle P^{-}P^{\prime0}|$, while $f_{+}^{P^{-}P^{\prime0}}(s)$ is the $P$-wave component, and they are known as the scalar and vector form factors accordingly.
Notice that the scalar contribution is suppressed by the mass-squared difference $\Delta_{P^{-}P^{\prime 0}}=m_{P^{-}}^{2}-m_{P^{\prime 0}}^{2}$.
In terms of these form factors, the differential decay width reads 
\begin{eqnarray} \label{spectral function}
& & \frac{d\Gamma\left(\tau^-\to P^{-}P^{\prime0}\nu_\tau\right)}{ds} = \frac{G_F^2M_\tau^3}{768\pi^3}S_{EW}|V_{{\rm{CKM}}}|^2\mathcal{C}_{P^{-}P^{\prime 0}}^{2}
\left(1-\frac{s}{M_\tau^2}\right)^2\nonumber\\
& & \left\lbrace\left(1+\frac{2s}{M_\tau^2}\right)\lambda_{P^{-}P^{\prime0}}^{3/2}(s)\big|f_{+}^{P^{-}P^{\prime0}}(s)\big|^2+\frac{3\Delta_{K\eta^{(\prime)}}^2}{s^{2}}\lambda_{P^{-}P^{\prime0}}^{1/2}(s)\big|f_{0}^{P^{-}P^{\prime0}}(s)\big|^2\right\rbrace\,,
\end{eqnarray}
where $\lambda_{P^{-}P^{\prime0}}\equiv\lambda(s,m_{P^{-}}^{2},m_{P^{0}}^{2})/s^{2}$ and $S_{\rm{EW}}$ is a short-distance electroweak correction.

Our initial setup approach to describe the required vector form factors assumes a Vector Meson Dominance form that includes both the real and imaginary parts of the unitary loop corrections thus fulfilling analyticity and unitarity.
One can then extract its phase $\phi^{P^{-}P^{0}}_{\rm{input}}(s)$ and insert it into a dispersion relation. 
The use of a thrice-subtracted dispersion relation
\begin{equation}
f_{+}^{P^{-}P^{\prime0}}(s)=\exp\left[\alpha_{1}s+\frac{\alpha_{2}}{2}s^{2}+\frac{s^{3}}{\pi}\int_{s_{\rm{th}}}^{\infty}ds^{\prime}\frac{\phi^{P^{-}P^{0}}_{\rm{input}}(s^{\prime})}{(s^{\prime})^{3}(s^{\prime}-s-i0)}\right]\,,
\label{FFthreesub}
\end{equation}
where $\alpha_{1,2}$ are two subtraction constants that can be related to chiral low-energy observables and $s_{\rm{th}}$ is the corresponding two-particle production threshold, is found to be an optimal choice that makes the fit less sensitive to the higher-energy region of the dispersive integral where the phase is less well-known.
In the isospin limit no scalar contributes to $\tau^{-}\to\pi^{-}\pi^{0}\nu_{\tau}$, while for the required $K\pi,K\eta^{(\prime)}$ scalar form factors, we use \cite{Jamin:2001zq}.

\subsection{The pion vector form factor and $\tau^{-}\to K^{-}K_{S}\nu_{\tau}$ decay}
\label{sec-2}

\begin{figure}[h]
\centering
\includegraphics[width=6.5cm,clip]{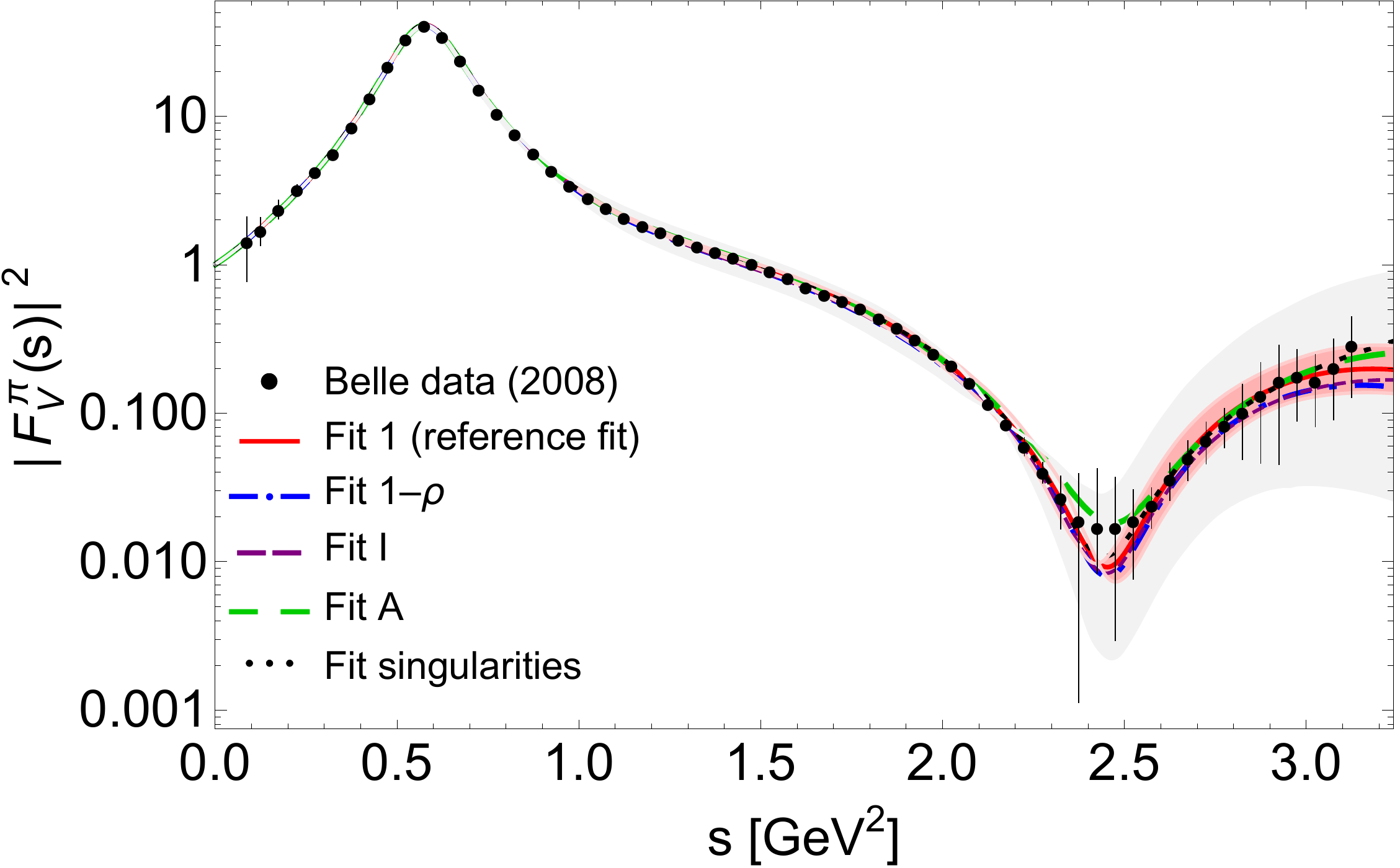}\includegraphics[width=6.5cm,clip]{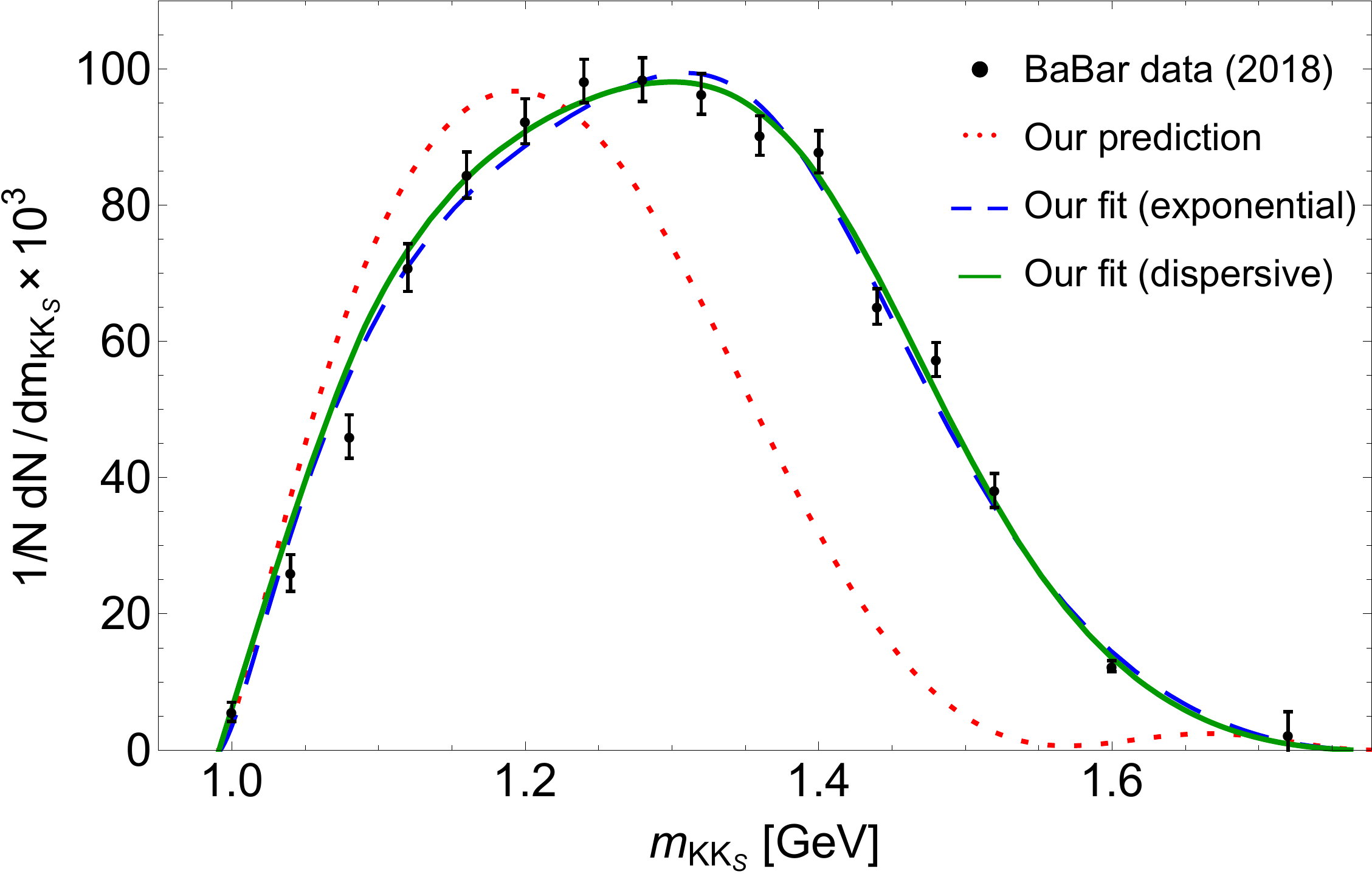}
\caption{Belle measurement of the modulus squared of the pion vector form factor \cite{Fujikawa:2008ma} (left) and BaBar data \cite{BaBar:2018qry} for $\tau^{-}\to K^{-}K_{S}\nu_{\tau}$ (right) as compared to our fits. See Ref.\,\cite{Gonzalez-Solis:2019iod} for details.}
\label{fig-2}       
\end{figure}

The pion vector form factor is a classic object in low-energy QCD since it provides a privileged laboratory to study the effects of $\pi\pi$ interactions under rather clean conditions.
In \cite{Gonzalez-Solis:2019iod}, we have exploited the synergy between Chiral Perturbation Theory and dispersion relations and provided a representation that uses for the phase required as input in Eq.\,(\ref{FFthreesub}):
\begin{equation}
\phi_{\rm{input}}^{\pi\pi}(s)=\left\{ \begin{array}{llll}
         \delta_{1}^{1}(s)&&&4m_{\pi}^{2}\le s<1\,\rm{GeV}^{2}\,,\\[1ex]
       \psi(s)&&& 1\,{\rm{GeV}}^{2}\le s<m_{\tau}^{2}\,.\\[1ex]
       \psi_{\infty}(s)&&&m_{\tau}^{2}\le s\,.\end{array} \right.
\label{PhaseRegions}       
\end{equation}
This phase contains the following remarkable features: $i)$ it fully exploits Watson's theorem providing a model-independent description of the elastic region i.e. until $\sim1$ GeV$^{2}$, through the use of the $\pi\pi$ scattering phase $\delta_{1}^{1}(s)$ \cite{GarciaMartin:2011cn}; $ii)$ for the region $m_{\tau}^{2}\leq s$, we guide smoothly the phase to $\pi$ at high-energies thus ensuring the correct $1/s$ fall-off; $iii)$ for the intermediate region $1\,{\rm{GeV}}^{2}\le s<m_{\tau}^{2}$, we use a parametrization that contains the physics of the inelastic regime until $m_{\tau}^{2}$ by means of $\psi(s)=\arctan[{\rm{Im}}f_{+}^{\pi\pi}(s)|^{3\,\rm{res}}_{\rm{expo}}/{\rm{Re}}f_{+}^{\pi\pi}(s)|^{3\,\rm{res}}_{\rm{expo}}]$, where $f_{+}^{\pi\pi}(s)|^{3\,\rm{res}}_{\rm{expo}}$ is the Omn\`{e}s exponential representation of the form factor that reads (see Ref.\,\cite{Gonzalez-Solis:2019iod} for details)
\begin{eqnarray}
f_{+}^{\pi\pi}(s)|^{3\,\rm{res}}_{\rm{expo}}&=&\frac{M_{\rho}^{2}+s\left(\gamma e^{i\phi_{1}}+\delta e^{i\phi_{2}}\right)}{M_{\rho}^{2}-s-iM_{\rho}\Gamma_{\rho}(s)}\exp\Bigg\lbrace {\rm{Re}}\Bigg[-\frac{s}{96\pi^{2}F_{\pi}^{2}}\left(A_{\pi}(s)+\frac{1}{2}A_{K}(s)\right)\Bigg]\Bigg\rbrace\nonumber\\[2mm]
&&-\gamma\frac{s\,e^{i\phi_{1}}}{M_{\rho^{\prime}}^{2}-s-iM_{\rho^{\prime}}\Gamma_{\rho^{\prime}}(s)}\exp\Bigg\lbrace-\frac{s\Gamma_{\rho^{\prime}}(M_{\rho^{\prime}}^{2})}{\pi M_{\rho^{\prime}}^{3}\sigma_{\pi}^{3}(M_{\rho^{\prime}}^{2})}{\rm{Re}}A_{\pi}(s)\Bigg\rbrace\nonumber\\[2mm]
&&-\delta\frac{s\,e^{i\phi_{2}}}{M_{\rho^{\prime\prime}}^{2}-s-iM_{\rho^{\prime\prime}}\Gamma_{\rho^{\prime\prime}}(s)}\exp\Bigg\lbrace-\frac{s\Gamma_{\rho^{\prime\prime}}(M_{\rho^{\prime\prime}}^{2})}{\pi M_{\rho^{\prime\prime}}^{3}\sigma_{\pi}^{3}(M_{\rho^{\prime\prime}}^{2})}{\rm{Re}}A_{\pi}(s)\Bigg\rbrace\,.
\label{FFExpThreeRes}
\end{eqnarray}
Armed with this parametrization, and variants of it, we have analyzed the high-statistics Belle data \cite{Fujikawa:2008ma} focusing our effort on the improvement of the description of the energy region where the $\rho(1450)$ and $\rho(1700)$ come up into play. 
In Fig.\,\ref{fig-2} (left), we display the form factor modulus squared including the statistical fit uncertainty for our reference fit (red error band) and a conservative systematic uncertainty coming from the largest variations of central values with respect to our reference fit (gray error band).
Our central results for the physical resonance mass and width of the three participating resonances are found to be \cite{Gonzalez-Solis:2019iod}
\begin{eqnarray}
& & M^{\rm{pole}}_{\rho}\,=\,760.6\pm0.8\,\,\rm{MeV}\,,\quad \Gamma^{\rm{pole}}_{\rho}\,=\,142.0\pm0.4\,\,\rm{MeV}\,,\nonumber\\[1mm]
& & M^{\rm{pole}}_{\rho^{\prime}}\,=\,1289\pm8^{+52}_{-143}\,\,\rm{MeV}\,,\quad \Gamma^{\rm{pole}}_{\rho^{\prime}}\,=\,540\pm16^{+151}_{-111}\,\,\rm{MeV}\,,\nonumber\\[1mm]
& & M^{\rm{pole}}_{\rho^{\prime\prime}}\,=\,1673\pm4^{+68}_{-125}\,\,\rm{MeV}\,,\quad \Gamma^{\rm{pole}}_{\rho^{\prime\prime}}\,=\,445\pm8^{+117}_{-49}\,\,\rm{MeV}\,,
\label{Polesfitpipi}
\end{eqnarray}
where the first error is statistical while the second is our estimated systematic uncertainty. 
From our study, we conclude that the determination of the pole mass and width of the $\rho(1450)$ and $\rho(1700)$ is limited by theoretical errors that have been usually underestimated so far.




%

\begin{thebibliography}{}
%
%

\bibitem{Gonzalez-Solis:2019iod}
  S.~Gonz\`{a}lez-Sol\'{i}s and P.~Roig,
  arXiv:1902.02273 [hep-ph].

\bibitem{Escribano:2014joa} 
  R.~Escribano, S.~Gonz\`{a}lez-Sol\'{i}s, M.~Jamin and P.~Roig,
  JHEP {\bf 1409}, 042 (2014)
  [arXiv:1407.6590 [hep-ph]].
  
\bibitem{Escribano:2013bca} 
  R.~Escribano, S.~Gonz\`{a}lez-Sol\'{i}s and P.~Roig,
  JHEP {\bf 1310}, 039 (2013)
  [arXiv:1307.7908 [hep-ph]].

\bibitem{Escribano:2016ntp} 
  R.~Escribano, S.~Gonz\`{a}lez-Sol\'{i}s and P.~Roig,
  Phys.\ Rev.\ D {\bf 94}, no. 3, 034008 (2016)
  [arXiv:1601.03989 [hep-ph]].

\bibitem{Jamin:2001zq}
  M.~Jamin, J.~A.~Oller and A.~Pich,
  Nucl.\ Phys.\ B {\bf 622} (2002) 279 [[hep-ph/0110193]].

\bibitem{GarciaMartin:2011cn} 
  R.~Garc\'{i}a-Mart\'{i}n, R.~Kaminski, J.~R.~Pel\'{a}ez, J.~Ruiz de Elvira and F.~J.~Yndur\'{a}in,
  Phys.\ Rev.\ D {\bf 83}, 074004 (2011)
  [arXiv:1102.2183 [hep-ph]].

\bibitem{Fujikawa:2008ma} 
  M.~Fujikawa {\it et al.} [Belle Collaboration],
  Phys.\ Rev.\ D {\bf 78}, 072006 (2008)
  [arXiv:0805.3773 [hep-ex]].

\bibitem{BaBar:2018qry} 
  J.~P.~Lees {\it et al.} [BaBar Collaboration],
  Phys.\ Rev.\ D {\bf 98}, no. 3, 032010 (2018)
  [arXiv:1806.10280 [hep-ex]].

\bibitem{Epifanov:2007rf}
D.~Epifanov {\it et al.} [Belle Collaboration],
Phys. Lett. B {\bf 654} (2007) 65.

\bibitem{Inami:2008ar}
  K.~Inami {\it et al.}  [Belle Collaboration],
  Phys.\ Lett.\ B {\bf 672} (2009) 209.

\bibitem{Boito:2008fq} 
  D.~R.~Boito, R.~Escribano and M.~Jamin,
  Eur.\ Phys.\ J.\ C {\bf 59}, 821 (2009)
  [arXiv:0807.4883 [hep-ph]].

\bibitem{Boito:2010me} 
  D.~R.~Boito, R.~Escribano and M.~Jamin,
  JHEP {\bf 1009}, 031 (2010)
  [arXiv:1007.1858 [hep-ph]].

\bibitem{Aubert:2007jh}
B.~Aubert {\it et al.} [BaBar Collaboration],
Phys. Rev. D {\bf 76} (2007) 051104.

\bibitem{Hayasaka:2009zz} 
  K.~Hayasaka [Belle Collaboration],
  PoS EPS {\bf -HEP2009}, 374 (2009).


\end{thebibliography}
%
%


The study of the $\tau^{-}\to K^{-}K_{S}\nu_{\tau}$ decay is of timely interest due to the recent measurement of its spectrum released by the BaBar Collaboration \cite{BaBar:2018qry}.
The $K^{-}K_{S}$ threshold opens around 1000 MeV which is $\sim$100 MeV larger than $M_{\rho}+\Gamma_{\rho}$, a characteristic energy scale for the $\rho(770)$-dominance region.
This implies that this mode is not sensitive to the $\rho(770)$ peak, and consequently not useful to study its properties, but rather enhances its sensitivity to the properties of the heavier copies $\rho(1450)$ and $\rho(1700)$.
In \cite{Gonzalez-Solis:2019iod}, within a dispersive parametrization of the kaon vector form factor, we have performed different fits to the measured spectrum (see right plot of Fig.\,\ref{fig-2}) and determined the $\rho(1450)$ mass and width.
We have pointed out that higher-quality data on this channel will allow to extract the $\rho(1450)$ and $\rho(1700)$ parameters with improved precision from a combined analysis with the pion vector form factor data.


\subsection{Combined analysis of the decays $\tau^{-}\to K_{S}\pi^{-}\nu_{\tau}$ and $\tau^{-}\to K^{-}\eta\nu_{\tau}$}
\label{sec-4}

We analyze the experimental measurement of the invariant mass distribution of the decay $\tau^{-}\to K_{S}\pi^{-}\nu_{\tau}$ together with spectrum of the $K^{-}\eta$ mode both released by Belle \cite{Epifanov:2007rf,Inami:2008ar}.
The former has been studied in detail in \cite{Boito:2008fq,Boito:2010me}, improving the determination of the resonance parameters of both the $K^{*}(892)$ and its first radial excitation $K^{*}(1410)$, while the later, with a threshold above the $K^{*}(892)$ dominance, has been studied in \cite{Escribano:2013bca} obtaining $K^{*}(1410)$ properties that are competitive with those of the $K_{S}\pi^{-}$ channel.
In \,\cite{Escribano:2014joa}, in a simultaneous study of the decay spectra of $\tau^{-}\to K_{S}\pi^{-}\nu_{\tau}$ and $\tau^{-}\to K^{-}\eta\nu_{\tau}$ within a dispersive representation of the required form factors, we have illustrated how the $K^{*}(1410)$ resonance parameters can be determined with improved precision as compared to previous studies.
We have also investigated possible isospin violations in the form factor slope parameters and claimed that making available the $K^{-}\pi^{0}$ decay spectrum \cite{Aubert:2007jh} would be extremely useful to get further insights.   

\begin{figure}[h]
\centering
\includegraphics[width=9.5cm,clip]{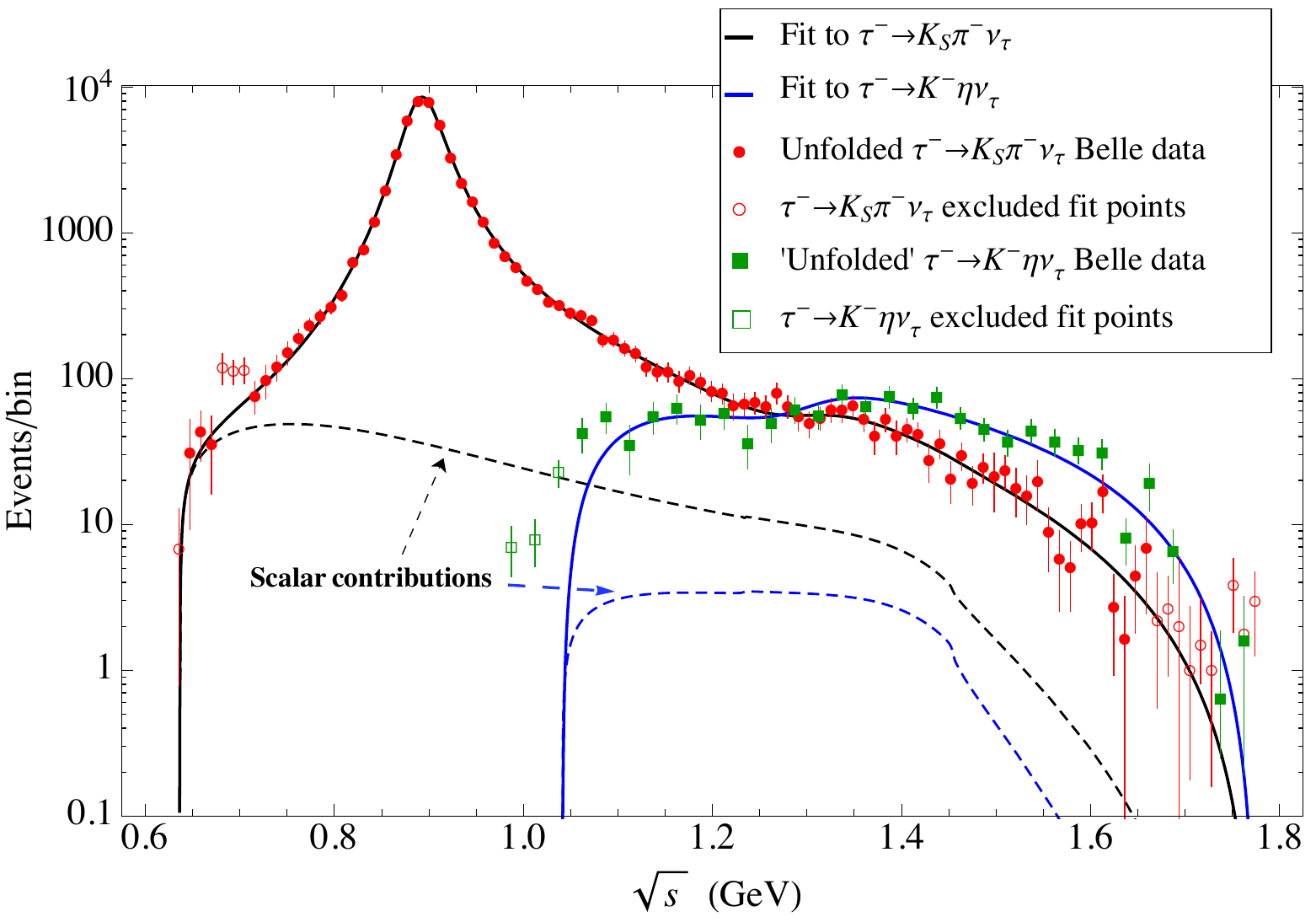}
\caption{Belle $\tau^{-}\to K_{S}\pi^{-}\nu_{\tau}$ (red circles) and $\tau^{-}\to K^{-}\eta\nu_{\tau}$ (green squares) measurements as compared to our best results (solid black and blue curves, respectively) obtained in combined fits to both data sets.}
\label{fig-3}       
\end{figure}

Our best fit results are compared to the measured Belle 
distributions in Fig.\,\ref{fig-3} where satisfactory agreement with data is seen for all data points.
The $K_{S}\pi^{-}$ decay channel is dominated by the $K^{*}(892)$ resonance peak followed by the contribution of the $K^{*}(1410)$ resonance, whose shoulder is visible on the second half of the spectrum.
The scalar form factor contribution is small although important to describe the data immediately above threshold.
There is no such clear peak structure for the $K\eta$ channel due to the interplay between both $K^{*}$ resonances.
The scalar form factor contribution is insignificant in this case.
With the current data, we succeed in improving the determination of the $K^{*}(1410)$ mass and width with the findings
\begin{equation}\label{K^*'}
 M_{K^{*}(1410)} \,=\, \left(1304 \pm 17\right)\,\mathrm{MeV} \,, \quad
\Gamma_{K^{*}(1410)} \,=\,\left(171 \pm 62\right)\,\mathrm{MeV}\,.
\end{equation}
For the $\tau^{-}\to K^{-}\eta^{\prime}\nu_{\tau}$ decay, it is dominated by the scalar form factor and we have obtained a branching ratio of $\sim1\times10^{-6}$ \cite{Escribano:2013bca}, well below the experimental upper bound. 

\subsection{The second-class current $\tau^{-}\to\pi^{-}\eta^{(\prime)}\nu_{\tau}$ decays}
\label{sec-5}
\begin{figure}[h]
\centering
\includegraphics[width=6.2cm,clip]{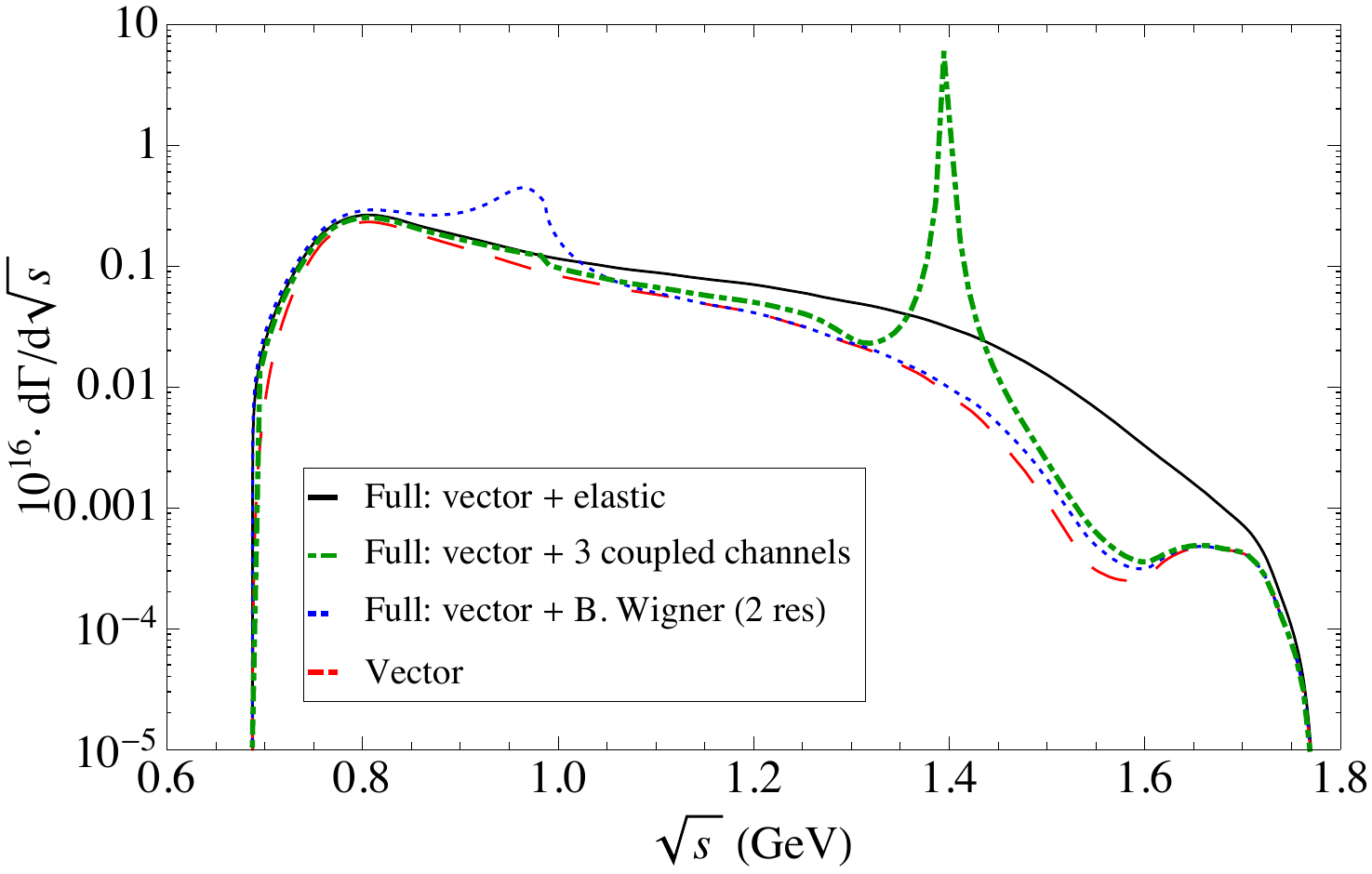} \includegraphics[width=6.3cm,clip]{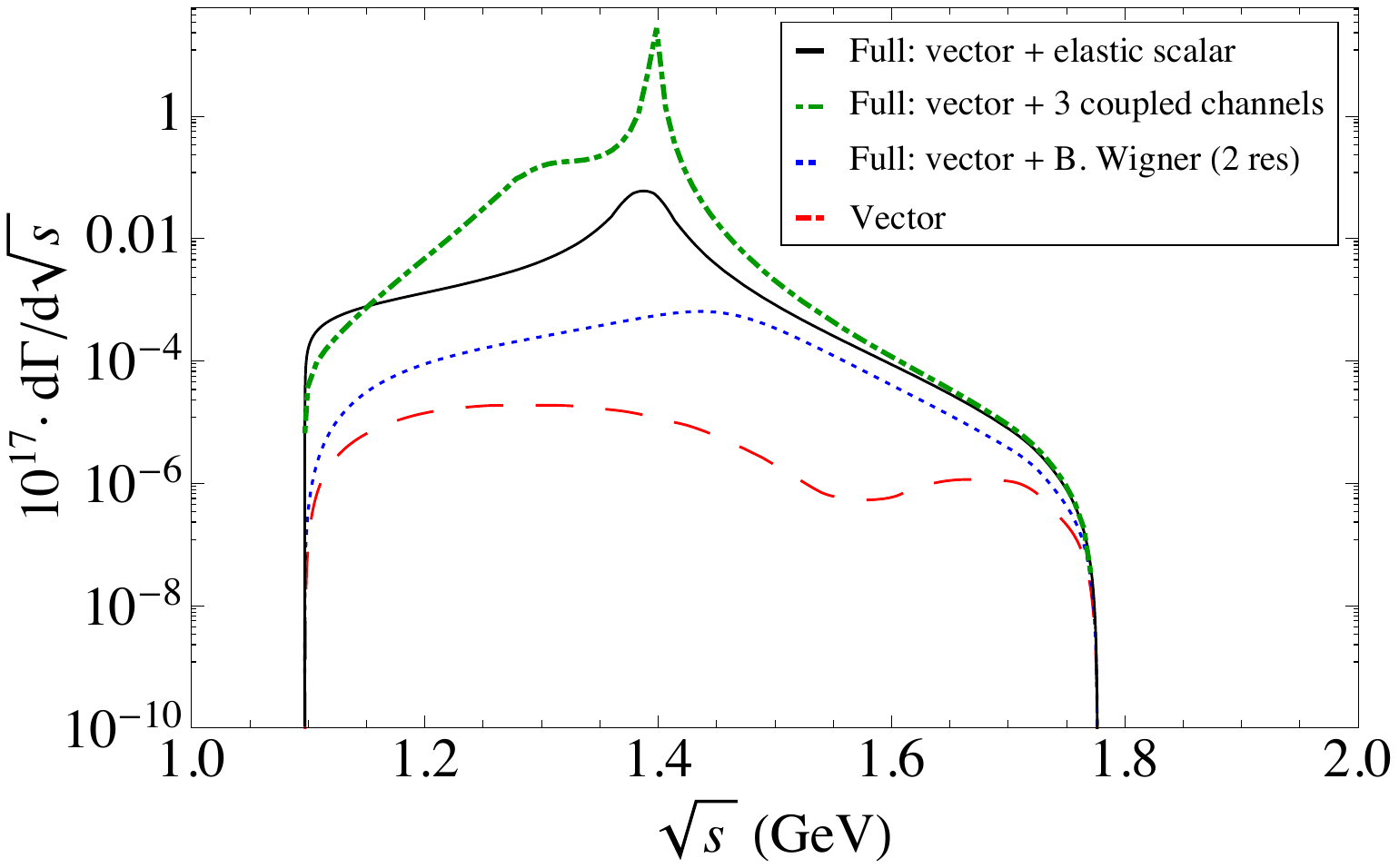}
\caption{Decay spectrum for $\tau^{-}\to\pi^{-}\eta\nu_{\tau}$ (left) and $\tau^{-}\to\pi^{-}\eta^{\prime}\nu_{\tau}$ (right).
See Ref.\,\cite{Escribano:2016ntp} for details.}
\label{fig-4}       
\end{figure}

The non-strange weak hadronic currents can be divided according to their $G$-parity: $i)$ first-class currents with quantum numbers $J^{PG}=0^{++},0^{--},1^{+-},1^{-+}$; $ii)$ second-class currents (SCC), which have $J^{PG}=0^{+-},0^{-+},1^{++},1^{--}$.
The former completely dominate weak interactions since there has been no evidence of the later in Nature so far.
We study the $\tau^{-}\to\pi^{-}\eta^{(\prime)}\nu_{\tau}$ decays which belong to the SCC processes i.e. parity conservation implies that these transitions must proceed through the vector current which has opposed $G$-parity to the $\pi^{-}\eta^{(\prime)}$ system. 
Our predictions \cite{Escribano:2016ntp} are displayed in
Fig.\,\ref{fig-4}, where we show the total decay rate distribution for $\tau^{-}\to\pi^{-}\eta\nu_{\tau}$ (left) and $\tau^{-}\to\pi^{-}\eta^{\prime}\nu_{\tau}$ (right).
The low-energy part of the $\pi\eta$ spectrum is dominated by the vector contribution associated to the $\rho(770)$ while effects of the $a_{0}(980)$ and $a_{0}(1450)$ scalar resonance contributions might show up and dominate the intermediate and high-energy part.
Contrarily, the vector contribution is suppressed in $\tau^{-}\to\pi^{-}\eta^{\prime}\nu_{\tau}$ because the $\pi^{-}\eta^{\prime}$ threshold lies well beyond the region of influence of the $\rho(770)$, thus being this mode dominated by the scalar form factor.
Our branching ratio predictions for $\pi^{-}\eta$ are found to be within the window $[0.36,2.12]\times10^{-5}$ respecting the current experimental upper limit, $7.3\times10^{-5}$ at $90\%$ CL, reported by Belle \cite{Hayasaka:2009zz}.
Regarding the branching of the $\pi^{-}\eta^{\prime}$ mode, it might be one or two order of magnitude smaller than the $\pi^{-}\eta$ channel.
 
\section{Summary}
\label{summary}
In this letter, we have provided an overview of all possible semileptonic two-meson decay channels of the $\tau$ lepton. 
These decays provide a privileged laboratory to study, under rather clean conditions, the energy region of two-meson form factors where resonances come up into play.
An ideal roadmap for describing them would require a model-independent approach demanding a full know\-led\-ge of QCD in both its perturbative and non-perturbative regimes, knowledge not yet unraveled. 
An alternative to such enterprise would pursuit a synergy between the formal theoretical calculations and experimental data.
In this respect, dispersion relations are a powerful tool to direct oneself towards a model-independent description of form factors.
By exploiting the synergy between dispersion relations and Chiral Perturbation Theory, we have carried out a dedicated study of the high-statistics Belle data of the pion vector form factor, assessing the role of the systematic uncertainties in the determination of the $\rho(1450)$ and $\rho(1700)$ parameters, and performed a first analysis of the $\tau^{-}\to K^{-}K_{S}\nu_{\tau}$ BaBar data.
We have also shown the potential of the combined analysis of $\tau^{-}\to K_{S}\pi^{-}\nu_{\tau}$ and $\tau^{-}\to K^{-}\eta\nu_{\tau}$ to extract the $K^{*}(1410)$ mass and width.
Finally, while for the decay $\tau^{-}\to\pi^{-}\eta\nu_{\tau}$ we find a total branching ratio that ranges $[0.36,2.12]\times10^{-5}$, well within the reach of Belle-II, the $\pi^{-}\eta^{\prime}$ channel might be one or two order of magnitude more suppressed.

\section*{Acknowledgements}
The author thanks the organizers of Phi-Psi 2019 for the very nice workshop we have enjoyed.
This work has been supported by the National Science Foundation (Grant No.\,PHY-1714253).

\end{document}